\documentclass[a4paper]{article}
\usepackage{ISCSLP2026}
\usepackage{ifthen}
\usepackage{multirow}
\usepackage{amssymb}
\usepackage{adjustbox}
\usepackage{booktabs}
\usepackage{enumitem}
\usepackage{bm}
\usepackage{siunitx}
\usepackage{makecell}
\usepackage{threeparttable}
\usepackage{url}

\newboolean{blind}
\setboolean{blind}{true} 
\title{StreamWSR: Streamable and Lightweight Waveform-Domain Neural Speech Super-Resolution}
\name{
    \ifthenelse{\boolean{blind}}
    {
        Yuan Tian,
        Yang Ai$^{*}$,
        Hui-Peng Du, Zhen-Hua Ling
        \thanks{\textnormal{$^{*}$ Corresponding author.
        This work was supported by the National Key Research and Development
        Program Project 2024YFE0217200 and the National Natural Science
        Foundation of China under Grant 62301521.}}
    }
    {Name$^1$, Co-author Name$^2$}
}
\address{
  \ifthenelse{\boolean{blind}}{National Engineering Research Center of Speech and Language Information Processing,\\
  University of Science and Technology of China, Hefei}
  {
  	$^1$Author Affiliation\\
  $^2$Co-author Affiliation
  }
}

\email{
	\ifthenelse{\boolean{blind}}{ytian1507@mail.ustc.edu.cn, yangai@ustc.edu.cn, redmist@mail.ustc.edu.cn, zhling@ustc.edu.cn}
	{author@university.edu, coauthor@company.com}
}

\begin{document}

\maketitle
\begin{abstract}

This paper proposes \textbf{StreamWSR}, a \textbf{Stream}able neural \textbf{W}aveform-domain model for speech \textbf{S}uper-\textbf{R}esolution (SR).
By adopting a fully causal architecture with compact frame-level waveform representation, the proposed StreamWSR supports zero-look-ahead streaming inference while avoiding vocoder-based reconstruction and explicit phase prediction.
Specifically, StreamWSR downsamples the input waveform into a compact frame-level representation using strided causal convolutions.
% In particular, StreamWSR first upsamples the low-resolution waveform to the target sampling rate and converts it into a compact frame-level sequence through causal convolution.
Then, a lightweight causal long-short-term modeling backbone is employed to capture both local waveform structures and long-range historical dependencies under causal constraints.
Finally, the modeled output is converted back to the waveform domain through a causal transposed-convolution and combined with the input waveform via a residual connection to generate the final high-resolution speech.
% Finally, the modeled output is upsampled by a causal transposed convolution to generate a waveform-domain residual, which is combined with the input waveform via a residual connection to generate the final high-resolution speech.
% Finally, the predicted residual sequence is converted back to the waveform domain through a causal transposed convolutional decoder and added to the upsampled input waveform to generate the high-resolution speech.
Experimental results on 16 kHz speech SR show that StreamWSR achieves competitive or superior speech quality and intelligibility compared with representative waveform- and spectrum-based baselines, while maintaining a zero-look-ahead streaming advantage with only 9M parameters and 2G FLOPs.
% These results demonstrate the effectiveness of StreamWSR for efficient zero-look-ahead streamable speech SR.

\end{abstract}

\section{Introduction}

Speech super-resolution (SR) aims to enhance low-resolution speech signals by reconstructing their missing high-frequency components and generating high-resolution speech.
By supplementing the effective high-frequency components of speech signals, speech SR can improve speech quality, intelligibility, and naturalness.
It has recently been explored in real-time communication \cite{li2021real} and codec-related scenarios \cite{buthe2025lightweight}, and has also been used to support downstream tasks such as speech enhancement and restoration \cite{andreev2023hifi++,liu2022voicefixer}, speech synthesis \cite{yang2025enhancing}, and automatic speech recognition \cite{li2019speech}.
In practical speech communication and interactive speech applications, speech SR models are expected not only to generate high-quality speech but also to operate with low algorithmic latency, where each output segment is generated using only the current and past input information without relying on future information.

% Early speech SR methods mainly relied on traditional statistical signal processing techniques \cite{chennoukh2001speech, 1326084}.
% However, these methods were limited by their insufficient modeling capacity and often suffered from over-smoothed reconstruction results \cite{7078992}.
% With the development of deep learning, neural network-based methods have achieved remarkable progress in speech SR by learning the mapping from low-resolution speech to high-resolution speech with powerful generation backbones.
Early statistical methods suffered from limited modeling capacity
\cite{chennoukh2001speech,1326084,7078992}, while recent neural
approaches have substantially improved speech SR performance.

Existing neural speech SR methods can generally be categorized into waveform-based and spectrum-based approaches according to their modeling targets.
Waveform-based methods directly generate high-resolution speech waveforms from low-resolution waveforms \cite{hao2020time,sui2024tramba}, preserving complete time-domain information and avoiding the information loss caused by intermediate acoustic representations.
% Compared with spectrum-domain representations, waveform-domain modeling preserves the sample-level temporal resolution of speech signals and avoids the information compression introduced by frame-level time-frequency analysis.
% However, this also leads to much longer sequences and makes efficient causal modeling more challenging.
%, especially under the zero-look-ahead setting.

To reduce the modeling difficulty of long waveform sequences, many speech SR methods perform reconstruction in the spectral domain. 
Mel-spectrogram-based methods first recover high-resolution mel spectrograms and then synthesize waveforms using neural vocoders \cite{yun2025flowhigh, shen2018natural}. 
This two-step strategy decomposes speech SR into a spectral reconstruction problem and a waveform generation problem, and has achieved promising results. 
For example, FLowHigh \cite{yun2025flowhigh} performs spectral-domain SR using single-step conditional flow matching, followed by vocoder-based waveform reconstruction. 
However, mel-spectrograms are compressed acoustic representations that discard phase information during feature extraction. 
Therefore, mel-spectrogram-based methods usually require an additional vocoder to implicitly reconstruct phase and generate the final waveform, which increases system complexity and may prevent fully end-to-end optimization \cite{liu2024audiosr, liu2022neural}.

Another line of spectral-domain methods adopts short-time Fourier transform (STFT) spectra as the modeling target. 
Compared with mel-spectrograms, STFT spectra provide more detailed time-frequency representations and can be inverted to waveforms when both amplitude and phase are available. 
However, phase modeling remains challenging due to its wrapped, highly nonlinear, and unstructured characteristics. 
As a result, many STFT-based methods mainly reconstruct amplitude spectra and estimate phase using heuristic signal processing techniques \cite{abel2018simple, gu2016speech}, which may limit the quality of the reconstructed speech. 
AP-BWE \cite{lu2024towards} alleviates this problem by explicitly predicting both amplitude and phase spectra with a dual-path network and reconstructing the waveform through inverse STFT (ISTFT). 
Nevertheless, explicit amplitude-phase prediction requires a more complex model structure and training objective. 
These limitations motivate a waveform-domain speech SR framework that preserves complete signal information while avoiding vocoder-based reconstruction and explicit phase prediction under zero-look-ahead low-latency constraints.

% For example, UDM+ \cite{yu2023conditioning} reconstructs high-frequency components through diffusion-based iterative denoising, while TRAMBA \cite{sui2024tramba} adopts a hybrid Transformer-Mamba architecture for efficient waveform-domain speech SR.
% However, waveform modeling may lead to much longer sequences and make efficient causal modeling more challenging.

% To reduce the modeling difficulty of long waveform sequences, many speech SR methods perform reconstruction in the spectral domain.
% Mel spectrogram-based methods first recover high-resolution mel spectrograms and then synthesize waveforms using neural vocoders \cite{yun2025flowhigh, shen2018natural}.
% This two-step strategy decomposes speech SR into simpler sub-problems and has achieved promising results.
% For instance, FLowHigh \cite{yun2025flowhigh} performs speech SR in the spectral domain using single-step conditional flow matching, followed by vocoder-based waveform reconstruction and post-processing.
% However, mel spectrograms are compressed acoustic representations that discard phase information during feature extraction.
% Therefore, mel spectrogram-based methods usually rely on an additional vocoder to implicitly reconstruct phase and generate the final waveform, which increases system complexity and may prevent fully end-to-end optimization \cite{liu2024audiosr, liu2022neural}.

\begin{figure*}[t]
	\centering
	\includegraphics[width=\textwidth]{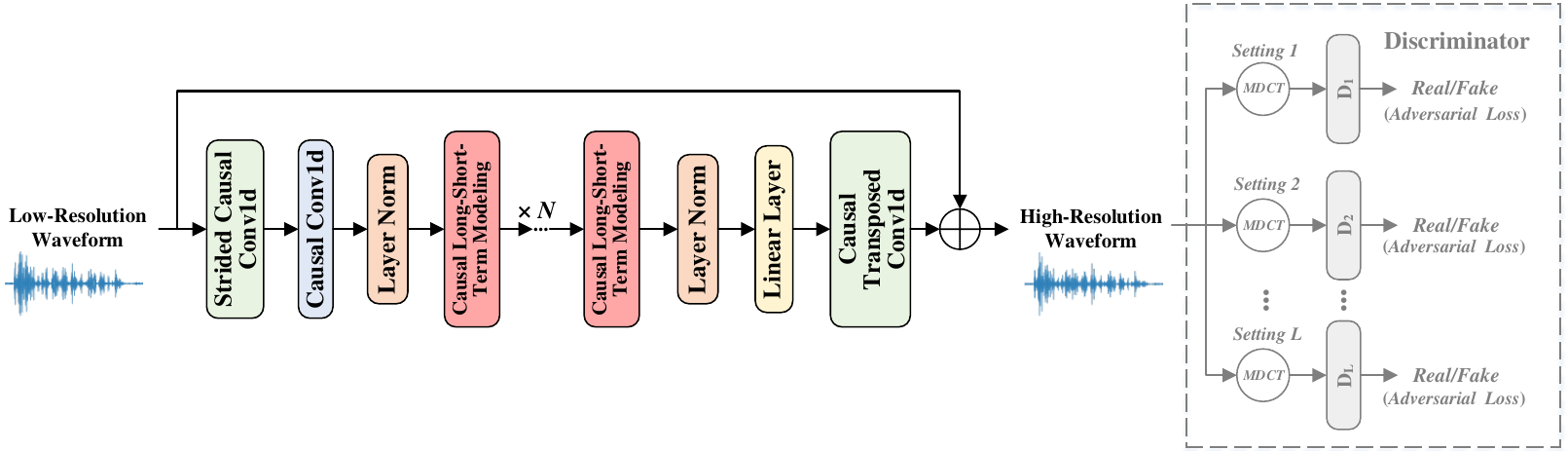}
	\caption{The overall structure of the proposed StreamWSR. The gray regions are appeared only during training. Here, \emph{Conv1d} represents the 1D convolution.}
 \label{fig1}
\end{figure*}

Therefore, we propose StreamWSR, a lightweight and fully causal waveform-domain speech SR model with compact frame-level representation. StreamWSR directly maps low-resolution waveforms to high-resolution waveforms in an end-to-end manner, while using spectral supervision only during training to improve high-frequency reconstruction and perceptual quality. Thus, StreamWSR enables zero-look-ahead streaming inference without vocoder-based reconstruction, explicit phase prediction, or additional inference cost. Experimental results on 16 kHz speech SR show that StreamWSR achieves competitive or superior speech quality and intelligibility compared with representative waveform- and spectrum-based baselines, while using only 9M parameters and 2G FLOPs.

%Aliquam quis orci consectetur nulla luctus ullamcorper. Suspendisse finibus luctus erat a dapibus.

\vspace{-1mm}

\section{Proposed Method}
\subsection{Model Structure}

The overview of the proposed StreamWSR is illustrated in Fig. \ref{fig1}.
Given a low-resolution waveform as input, StreamWSR aims to directly predict its high-resolution waveform in the time domain, forming an end-to-end waveform-domain speech SR framework.
The input low-resolution waveform is first upsampled to the target sampling rate using sinc interpolation, resulting in a high-frequency-depleted waveform.
StreamWSR then recovers the missing high-frequency components from this waveform under strict real-time constraints.

First, StreamWSR uses a strided causal convolution followed by a standard causal convolution to convert the upsampled waveform into a compact frame-level representation, which reduces the temporal resolution while preserving causal waveform-domain information.
The sequence is then processed by $N$ stacked causal long-short-term modeling blocks, inspired by ConvNeXt \cite{liu2022convnet} and attention mechanisms \cite{vaswani2017attention}.
Each block is designed to model both local and long-range historical dependencies in a causal manner.
For local historical modeling, a causal dilated convolution is used to enlarge the receptive field without accessing future samples, followed by point-wise convolutions and SnakeBeta activation \cite{ziyin2020neural} to enhance nonlinear representation.
For long-range historical modeling, masked multi-head self-attention is employed, where an upper-triangular mask prevents each frame from attending to future positions.
Residual connections and layer normalization \cite{ba2016layer} are applied throughout the block to stabilize feature transformation and preserve temporal information.

After the backbone, a linear layer transforms the feature dimension, and the resulting representation is converted into a waveform-domain residual through a causal transposed-convolution.
% After the backbone, a linear layer predicts residual tokens, which are converted by a causal transposed convolutional decoder into a waveform-domain residual.
Finally, the predicted residual is added to the upsampled low-resolution waveform to generate the high-resolution waveform.
Since all convolutional and attention operations are causal, StreamWSR supports zero-look-ahead streamable inference while keeping the generation pipeline fully in the waveform domain.

\vspace{-1mm}
\subsection{Training Strategies}

To improve the perceptual quality and spectral fidelity of the generated high-resolution speech, StreamWSR is trained with a spectrally guided adversarial framework based on generative adversarial networks (GANs) \cite{goodfellow2014generative}.
It is worth noting that the generator itself operates directly in the waveform domain, taking the low-resolution waveform as input and producing the predicted high-resolution waveform as output.
The spectral representations introduced in this section are used only for training-time supervision and do not introduce any additional computational cost during inference.

\vspace{-1mm}
\subsubsection{Multi-resolution Spectral Adversarial Training}

Although StreamWSR generates speech waveforms directly, we employ a multi-resolution spectral discriminator to provide structured time-frequency supervision.
Given the predicted high-resolution waveform and the natural high-resolution waveform, the discriminator extracts their modified discrete cosine transform (MDCT) spectra using multiple MDCT configurations.
Each configuration corresponds to one sub-discriminator and provides a different time-frequency resolution, allowing the discriminator to evaluate the generated speech from multiple spectral scales.
As shown in Fig. \ref{fig1}, the discriminator consists of $L$ parallel sub-discriminators $D_1,D_2,\dots,D_L$.
Each sub-discriminator processes the MDCT spectrum extracted under a specific configuration and produces a real or fake score.
Specifically, each sub-discriminator is implemented by several cascaded 2D convolutional blocks with LeakyReLU activations \cite{maas2013rectifier}, followed by a single-channel convolutional layer for score prediction.
The generator and discriminator are optimized using a hinge adversarial objective. 
In addition, a feature matching loss $\mathcal{L}_{FM}$ \cite{kumar2019melgan} is applied to intermediate discriminator features to stabilize adversarial training and improve perceptual consistency. 
% In addition, a feature matching loss $\mathcal{L}_{FM}$ \cite{kumar2019melgan} is applied between the intermediate discriminator features of natural and generated speech, which helps stabilize adversarial training and improves the perceptual consistency of the generated waveform.

\setlength{\tabcolsep}{2pt} % 默认一般是 6pt，调小能显著减宽
\begin{table*}[t!]
\renewcommand{\arraystretch}{1}
	\centering
	\caption{The speech quality evaluation results of the proposed StreamWSR and speech SR baselines evaluated on the test set of the VCTK dataset with a target sampling rate of 16 kHz. The \textbf{bold} and \underline{underline} numbers indicate the optimal and sub-optimal results, respectively.}\label{tab1}
	%\adjustbox{width=\textwidth}{
    %\resizebox{\textwidth}{!}{
		\begin{tabular}{l c c | c c c c c c | c c  }
			\specialrule{1.2pt}{0pt}{0pt}
			\multirow{2}{*}{Methods} & \multirow{2}{*}{Type} & \multirow{2}{*}{Streamable} &\multicolumn{2}{c}{8 kHz$\rightarrow$16 kHz} & \multicolumn{2}{c}{4 kHz$\rightarrow$16 kHz} & \multicolumn{2}{c|}{2 kHz$\rightarrow$16 kHz}  & \multirow{2}{*}{Params.$\downarrow$} & \multirow{2}{*}{FLOPs$\downarrow$} \\
              & && {LSD$\downarrow$} & {ViSQOL$\uparrow$} &  {LSD$\downarrow$} & {ViSQOL$\uparrow$} &   {LSD$\downarrow$} & {ViSQOL$\uparrow$}   \\
			\hline
			
            {UDM+} & waveform & $\times$ &0.88&4.57 & 1.16&3.99   & 1.33 & 3.35 & \underline{6.3 M} & 189.54 G \\
            
            {TRAMBA} & waveform & $\times$ & 0.79  & 4.64   & 0.97 & 4.25  & 1.07 & 3.62 & \textbf{5.18 M} & \textbf{0.72 G} \\
            
            % {NVSR} & spectrum & $\times$ &0.79&4.52   &0.95&4.11    & 1.1 & 3.41 \\
            
            {FLowHigh} & spectrum & $\times$ & 1.22 & \underline{4.70}  &   1.30 & \textbf{4.31}    & 1.62 & 2.98  & 49.4 M & 212.93 G\\
            {AP-BWE} & spectrum & $\times$ &\textbf{0.69}& \textbf{4.71}   &\textbf{0.87} & \underline{4.30} & \textbf{0.99} & \underline{3.76} & 29.8 M & 5.97 G  \\
            
       		% {mdctGAN} & spectrum & $\times$ &0.83& 4.54    &1.05 & 3.98   &1.24 &3.22  \\
		    {StreamWSR} & waveform & $\checkmark$ & \underline{0.73} & 4.68 & \underline{0.92} & 4.27 & \underline{1.03} & \textbf{3.81} & 9.03 M & \underline{2.12 G} \\
        
			%\hline
        
			%\hline
			
			\specialrule{1.2pt}{0pt}{0pt}
	\end{tabular}%}
\end{table*}

\vspace{-1mm}
\subsubsection{Spectral Reconstruction Losses}

In addition to adversarial training, we introduce two spectral reconstruction losses to further constrain the generated high-resolution waveform.
First, an MDCT spectral loss $\mathcal{L}_{FW\text{-}MDCT}$ is used to measure the reconstruction error between the MDCT spectra of the generated and natural high-resolution speech.
A simple frequency-weighting strategy is applied to this loss, where higher frequency bins are assigned slightly larger weights to encourage the recovery of missing high-frequency components.
Second, a mel-spectrogram loss $\mathcal{L}_{Mel}$ is employed to provide perceptually related spectral supervision.
It is computed from both L1 and L2 distances between the mel-spectrograms extracted from the generated and natural high-resolution waveforms.

Therefore, the overall generator loss is defined as: 
\begin{equation} 
\mathcal{L}_{G} = \mathcal{L}_{adv} + \mathcal{L}_{FM} + \lambda_{FW\text{-}MDCT}\mathcal{L}_{FW\text{-}MDCT} + \lambda_{Mel}\mathcal{L}_{Mel}, \label{eq:overall_loss} 
\end{equation} 
where, $\mathcal{L}_{adv}$ denotes the adversarial loss.
$\lambda_{FW\text{-}MDCT}$ and $\lambda_{Mel}$ are hyperparameters used to balance the spectral reconstruction losses.
During training, the StreamWSR and its discriminator are alternately optimized using the generator loss and the discriminator adversarial loss, respectively.

\begin{table*}[t!]
\renewcommand{\arraystretch}{1}
\centering
\caption{The intelligibility evaluation results of the proposed StreamWSR and speech SR baselines evaluated on the test set of the VCTK dataset with a target sampling rate of 16 kHz. The \textbf{bold} and \underline{underline} numbers indicate the optimal and sub-optimal results, respectively.}
\label{tab2}
%\resizebox{\columnwidth}{!}{%
    \begin{tabular}{l | c c c c c c c c c c }
        \specialrule{1.2pt}{0pt}{0pt}
         \multirow{2}{*}{Methods} & \multirow{2}{*}{Streamable} & \multicolumn{3}{c}{8 kHz$\rightarrow$16 kHz} & \multicolumn{3}{c}{4 kHz$\rightarrow$16 kHz} & \multicolumn{3}{c}{2 kHz$\rightarrow$16 kHz} \\
         & & {WER(\%)$\downarrow$} & {CER(\%)$\downarrow$} & {STOI(\%)$\uparrow$} & {WER(\%)$\downarrow$} &{CER(\%)$\downarrow$} & {STOI(\%)$\uparrow$} & {WER(\%)$\downarrow$} & {CER(\%)$\downarrow$} & {STOI(\%)$\uparrow$}\\
        \hline
        {UDM+} & $\times$ & 4.50&2.16& \underline{99.70} & 20.37 & 13.00 & 91.95 & 86.64 & 67.65 & 81.92 \\
        
        {TRAMBA} & $\times$ &  \underline{3.73}  & \underline{1.68} & {99.47}  & {10.27} & {5.86} & {94.32}  & \textbf{33.31} & \textbf{23.57}  & \textbf{87.68}  \\
        
        % {NVSR} & $\times$ & 4.38 & 2.02 & 98.84 & 13.56 & 8.51 & 92.04 & 59.53 & 44.43 & 82.38  \\

        {FLowHigh} & $\times$ & 3.81 & 1.75 & 99.66 & \underline{9.49} & \underline{5.39} & \textbf{94.96} & 60.80 & 46.42 & 78.59 \\
        
        {AP-BWE} & $\times$ & \textbf{3.72}& \textbf{1.67} & \textbf{99.77} & \textbf{6.69} & \textbf{3.54} &    \underline{94.75} &  \underline{36.69} & \underline{25.61} & {87.00}  \\
        
        % {mdctGAN} & $\times$ & 4.34& 1.99 & 99.33 & 11.30 & 6.82 &90.00 & 54.69 & 40.55 & 76.60 \\    
        {StreamWSR} & $\checkmark$ & 3.84 & {1.79} & \textbf{99.77} & 9.85 & {5.55} & {94.30} & {39.33} & {27.49} & \underline{87.14}   \\

% {TRAMBA} &  3.82  & 1.71 & 99.67 & 7.79 & 4.39 &  94.91 & 33.58 & 23.39  & 88.03  \\  
        \specialrule{1.2pt}{0pt}{0pt}
    \end{tabular}%
%}
\end{table*}

\begin{table}[t]
\renewcommand{\arraystretch}{1.2}
	\centering
	\caption{Ablation study results of the proposed StreamWSR under the 2 kHz$\rightarrow$16 kHz SR setting.}\label{tab3}
	\adjustbox{width=\linewidth}{
		\begin{tabular}{c| c c c c c c c }
        \specialrule{1.2pt}{0pt}{0pt}
			{}& {LSD$\downarrow$} & {ViSQOL$\uparrow$} & {WER$\downarrow$} & {CER$\downarrow$} & {STOI$\uparrow$} & {Params.$\downarrow$} & {FLOPs$\downarrow$} \\
			\hline
            % \times & \times & \times & 1.04 & 3.79 & 36.57 & 25.76 & 87.08 & 0.86 & 3.32 & 3.74 & 1.76  & \textbf{99.80} \\
            
             % replace mrd & 1.05 & 3.78 & 43.49 & 30.84  & 86.71 \\
            StreamWSR & 1.03 & 3.81 & 39.33 & 27.49  & 87.14 & 9.03 M & 2.12 G\\
             rep. D & 1.05 & 3.78 & 43.49 & 30.84  & 86.71 & 9.03 M & 2.12 G \\
            % 3 &  &  fake &  1.30 & 3.83 & 39.79 & 28.14 & 87.73 \\
            % fake up-down & 1.07 & 3.69 & 39.46 & 28.48 & 87.16 \\
            w/o Stride & 1.07 & 3.69 & 39.46 & 28.48 & 87.16 &  8.92 M & 78.4 G \\
            % 3&$\checkmark$  & $\checkmark$ & 1.01 & 3.76 & 43.97 & 31.21 \\     
            % \checkmark &\times& \times & 1.06 & 3.76 & 38.64 & 27.33 & 86.36 & 0.74 & 3.31 & 3.76 & 1.70 & 99.50\\
  			% 4&$\checkmark$ &$\checkmark$& $\times$ & 1.08 & 3.77 & 37.04 & 25.76 & 87.48  \\
		   %  5& {$\checkmark$} & $\checkmark$ & $\checkmark$ & 1.04 & 3.82 & 36.27 & 25.25 & 87.67  \\
			%\hline
			
\specialrule{1.2pt}{0pt}{0pt}
	\end{tabular}}
\end{table}

\vspace{-1mm}
\section{Experiments and Results}
\subsection{Experimental Setup}
\subsubsection{Dataset}

In our experiments\footnote{Speech samples are available at: \url{https://tian1507.github.io/StreamWSR/}.}, we used the VCTK-0.92 dataset \cite{yamagishi2019cstr}, which contains approximately 44 hours of 48 kHz speech from 110 English speakers with various accents.
% Following previous speech SR studies \cite{liu2022neural}, we used only mic1 recordings and excluded speakers p280 and p315.    删了
All recordings were downsampled to 16 kHz as high-resolution references. 
We considered three SR settings, where low-resolution inputs were generated by downsampling the 16 kHz waveforms to 8 kHz, 4 kHz, and 2 kHz, respectively, corresponding to extension factors of 2, 4, and 8.

%, and the corresponding low-resolution inputs were generated by further downsampling and upsampling back to 16 kHz.

\subsubsection{Implementation}

StreamWSR first used a strided causal 1D convolution with a kernel size of 80, a stride of 80, and 40 output channels to downsample the input waveform into a compact frame-level representation. 
The representation was then projected to 352 channels by a standard causal 1D convolution with a kernel size of 7. 
The backbone consisted of 8 causal long-short-term modeling blocks (i.e., $N=8$). 
In each block, the causal dilated depth-wise convolution used a kernel size of 7, a dilation factor of 2, and 352 channels. 
The hidden channel size of the point-wise transformation was set to 512, and the masked multi-head self-attention used 8 heads with an embedding dimension of 352. 
After the backbone, a linear layer transformed the feature dimension to 40, followed by a causal transposed 1D convolution with a kernel size of 80 and a stride of 80 to generate the waveform-domain residual.

For adversarial training, the multi-resolution spectral discriminator used three MDCT configurations (i.e., $L=3$), i.e., $(100,50,50)$, $(400,200,200)$, and $(40,20,20)$, where each tuple denotes the frame length, frame shift, and number of frequency bins. 
The MDCT spectral loss was computed using an MDCT configuration of $(80,40,40)$. 
The mel-spectrogram loss was computed using 80 mel filters, with an FFT size of 1024, a window size of 320, and a hop size of 40. 
During training, each waveform was randomly cropped into 16,000-sample segments. 
The model was optimized using AdamW with $\beta_1=0.8$, $\beta_2=0.99$, an initial learning rate of $2\times10^{-4}$, and an exponential decay factor of 0.999. 
The total number of training steps was 600k with a batch size of 16, and all experiments were conducted on a single NVIDIA RTX 3090 GPU.

\vspace{-1mm}

\subsubsection{Baselines}

We selected four representative neural speech SR methods as baselines, including waveform-based UDM+ \cite{yu2023conditioning} and TRAMBA \cite{sui2024tramba}, as well as spectrum-based FLowHigh \cite{yun2025flowhigh} and AP-BWE \cite{lu2024towards}.
UDM+ reconstructs high-frequency components through diffusion-based iterative denoising, while TRAMBA adopts a hybrid Transformer-Mamba architecture for efficient waveform-domain SR.
FLowHigh performs speech SR in the spectral domain using single-step conditional flow matching, followed by vocoder-based waveform reconstruction and post-processing.
AP-BWE predicts amplitude and phase spectra with a dual-path network and reconstructs the high-resolution waveform through ISTFT.
For fairness, all baseline models were trained and evaluated on the same dataset.
% For 16 kHz speech SR, we considered extension factors of 2, 4, and 8, corresponding to 8 kHz, 4 kHz, and 2 kHz inputs extended to 16 kHz, respectively.

\vspace{-1mm}
\subsubsection{Evaluation Metrics}

We evaluated the compared speech SR methods from three aspects, i.e., speech quality, intelligibility, and complexity.
For speech quality, log-spectral distance (LSD) and virtual speech quality objective listener (ViSQOL) \cite{chinen2020visqol} were used to measure spectral distortion and perceptual quality, respectively.
For intelligibility, we used Whisper \cite{radford2023robust} to transcribe the extended speech and calculated word error rate (WER) and character error rate (CER), together with short-time objective intelligibility (STOI). 
For complexity, we reported the number of model parameters (Params.) to measure model complexity and the floating-point operations (FLOPs) required to generate one-second 16 kHz speech to measure computational complexity.
% For complexity, the number of model parameters (Params.) and floating point operations (FLOPs) required to generate one-second 16 kHz speech were reported.
% The speech quality and model/computational complexity results are summarized in Table \ref{tab1}, while the intelligibility evaluation results are presented in Table \ref{tab2}.
The speech quality and complexity results are summarized in Table \ref{tab1}, and the intelligibility results are reported in Table \ref{tab2}.

\vspace{-2mm}
\subsection{Results and Analysis}

\subsubsection{Comparison with Baseline Speech SR Methods}

% The objective results of speech quality, intelligibility, and model complexity are shown in Tables \ref{tab1} and \ref{tab2}.
As shown in Tables \ref{tab1} and \ref{tab2}, compared with UDM+, StreamWSR achieved clearly lower LSD and higher ViSQOL scores under all three SR settings, indicating more accurate spectral reconstruction and better perceptual quality.
In terms of intelligibility, UDM+ suffered from severe degradation as the extension factor increased, while StreamWSR maintained more stable results.
Although UDM+ had fewer parameters, its diffusion-based iterative sampling led to extremely high computational cost.
Compared with TRAMBA, StreamWSR achieved better speech quality across all extension factors.
This indicates that the proposed causal waveform-domain model can recover high-frequency components with better perceptual quality.
For intelligibility, StreamWSR achieved a level comparable to TRAMBA across different SR settings.
Although TRAMBA had fewer parameters and lower FLOPs, the quality of its generated speech was inferior to that of StreamWSR.
In addition, TRAMBA was not designed for zero-look-ahead streamable inference, whereas StreamWSR supported zero-look-ahead streamable inference, making it more suitable for strict low-latency scenarios. 
Compared with FLowHigh, StreamWSR achieved substantially lower LSD under all SR settings, showing better spectral reconstruction accuracy.
In terms of ViSQOL, StreamWSR achieved comparable perceptual quality to FLowHigh and even performed better under the extremely high extension-factor setting.
For intelligibility, FLowHigh showed a clear degradation under the extremely high extension-factor setting, while StreamWSR maintained better intelligibility.
In terms of complexity, FLowHigh required more parameters and FLOPs due to its flow-matching generation and vocoder-based reconstruction pipeline, whereas StreamWSR was much more lightweight and computationally efficient.
Compared with AP-BWE, StreamWSR achieved comparable speech quality and intelligibility overall.
However, AP-BWE relied on a complex dual-path amplitude-phase prediction framework and was not streamable.
In contrast, StreamWSR adopted a simpler waveform-domain architecture with fewer parameters and lower computational cost, while supporting zero-look-ahead streamable inference.
Overall, StreamWSR achieved a favorable balance among speech quality, intelligibility, complexity, and streamability.

\subsubsection{Ablation Studies}

To further analyze the effectiveness of the key designs in StreamWSR, we conducted ablation studies under the most challenging 2 kHz$\rightarrow$16 kHz SR setting, as shown in Table \ref{tab3}.
First, we replaced the multi-resolution spectral discriminator with a waveform-domain discriminator (denoted as rep. D), while keeping the generator and other training objectives unchanged.
Compared with StreamWSR, this variant led to degradation in both speech quality and intelligibility.
This indicates that, although the generator operates directly in the waveform domain, MDCT-domain adversarial supervision provides more effective structured time-frequency guidance for high-frequency reconstruction than waveform-domain discrimination.
Second, we removed the strided causal convolution and causal transposed-
convolution (denoted as w/o Stride).
Specifically, the initial strided causal convolution and the final causal transposed convolution were removed, and the backbone directly processed the full-length waveform sequence.
This variant obtained worse LSD and ViSQOL than StreamWSR, indicating that directly modeling the full-resolution waveform sequence makes it more difficult to recover high-quality spectral and perceptual details.
Although its intelligibility metrics were close to those of StreamWSR, its computational cost increased dramatically, with FLOPs reaching 78.4 G.
This demonstrates that compact frame-level waveform representation is crucial for reducing the temporal modeling burden of raw waveforms while maintaining effective speech reconstruction.
Overall, these comparative experiments show that the multi-resolution spectral discriminator and strided convolutional modules are both important for StreamWSR.
% The former provided effective spectral-domain adversarial guidance during training, while the latter reduced the temporal modeling burden of raw waveforms and improved the quality of streamable waveform-domain reconstruction.
The former provided effective spectral-domain adversarial guidance, while the latter reduced the temporal modeling burden and improved streamable waveform reconstruction.

\vspace{-1mm}

\section{Conclusion}

In this paper, we introduced StreamWSR, a streamable waveform-domain neural model for speech SR.
StreamWSR directly maps low-resolution waveforms to high-resolution waveforms with a fully causal architecture, avoiding vocoder-based reconstruction and explicit phase prediction.
By using compact frame-level waveform representation and causal long-short-term modeling, StreamWSR effectively captures both local waveform structures and long-range historical dependencies under zero-look-ahead constraints.
Experimental results on 16 kHz speech SR show that StreamWSR achieves competitive or superior speech quality and intelligibility compared with representative waveform- and spectrum-based baselines, while using only 9M parameters and 2G FLOPs.
Ablation experiments further confirm the effectiveness of the multi-resolution spectral discriminator and compact frame-level waveform representation.
In future work, we will improve the robustness and generalization ability of StreamWSR and explore its application to more real-time speech and audio tasks.

\bibliographystyle{IEEEtran}

\bibliography{mybib}

@article{ba2016layer,
  title={Layer normalization},
  author={Ba, Jimmy Lei and Kiros, Jamie Ryan and Hinton, Geoffrey E},
  journal={arXiv preprint arXiv:1607.06450},
  year={2016}
}

@inproceedings{liu2022convnet,
  title={A convnet for the 2020s},
  author={Liu, Zhuang and Mao, Hanzi and Wu, Chao-Yuan and Feichtenhofer, Christoph and Darrell, Trevor and Xie, Saining},
  booktitle={Proc. CVPR},
  pages={11976--11986},
  year={2022}
}

@article{vaswani2017attention,
  title={Attention is all you need},
  author={Vaswani, Ashish and Shazeer, Noam and Parmar, Niki and Uszkoreit, Jakob and Jones, Llion and Gomez, Aidan N and Kaiser, {\L}ukasz and Polosukhin, Illia},
  journal={Advances in neural information processing systems},
  volume={30},
  year={2017}
}

@article{ziyin2020neural,
  title={Neural networks fail to learn periodic functions and how to fix it},
  author={Ziyin, Liu and Hartwig, Tilman and Ueda, Masahito},
  journal={Advances in Neural Information Processing Systems},
  volume={33},
  pages={1583--1594},
  year={2020}
}

@article{goodfellow2014generative,
  title={Generative adversarial nets},
  author={Goodfellow, Ian J and Pouget-Abadie, Jean and Mirza, Mehdi and Xu, Bing and Warde-Farley, David and Ozair, Sherjil and Courville, Aaron and Bengio, Yoshua},
  journal={Advances in neural information processing systems},
  volume={27},
  year={2014}
}

@inproceedings{maas2013rectifier,
  title={Rectifier nonlinearities improve neural network acoustic models},
  author={Maas, Andrew L and Hannun, Awni Y and Ng, Andrew Y},
  booktitle={Proc. ICML},
  volume={30},
  number={1},
  pages={3},
  year={2013}
}

@article{kumar2019melgan,
  title={Melgan: Generative adversarial networks for conditional waveform synthesis},
  author={Kumar, Kundan and Kumar, Rithesh and De Boissiere, Thibault and Gestin, Lucas and Teoh, Wei Zhen and Sotelo, Jose and De Brebisson, Alexandre and Bengio, Yoshua and Courville, Aaron C},
  journal={Advances in neural information processing systems},
  volume={32},
  year={2019}
}

@article{yamagishi2019cstr,
  title={{CSTR VCTK} Corpus: English multi-speaker corpus for {CSTR} voice cloning toolkit (version 0.92)},
  author={Yamagishi, Junichi and Veaux, Christophe and MacDonald, Kirsten and others},
  journal={University of Edinburgh. The Centre for Speech Technology Research (CSTR)},
  pages={271--350},
  year={2019}
}

@inproceedings{yu2023conditioning,
  title={Conditioning and sampling in variational diffusion models for speech super-resolution},
  author={Yu, Chin-Yun and Yeh, Sung-Lin and Fazekas, Gy{\"o}rgy and Tang, Hao},
  booktitle={Proc. ICASSP},
  pages={1--5},
  year={2023}
}

@article{sui2024tramba,
  title={T{RAMBA}: A hybrid transformer and mamba architecture for practical audio and bone conduction speech super resolution and enhancement on mobile and wearable platforms},
  author={Sui, Yueyuan and Zhao, Minghui and Xia, Junxi and Jiang, Xiaofan and Xia, Stephen},
  journal={Proc. ACM IMWUT},
  volume={8},
  number={4},
  pages={1--29},
  year={2024}
}

@inproceedings{yun2025flowhigh,
  title={Flowhigh: Towards efficient and high-quality audio super-resolution with single-step flow matching},
  author={Yun, Jun-Hak and Kim, Seung-Bin and Lee, Seong-Whan},
  booktitle={Proc. ICASSP},
  pages={1--5},
  year={2025}
}

@article{lu2024towards,
  title={Towards high-quality and efficient speech bandwidth extension with parallel amplitude and phase prediction},
  author={Lu, Ye-Xin and Ai, Yang and Du, Hui-Peng and Ling, Zhen-Hua},
  journal={IEEE/ACM Transactions on Audio, Speech, and Language Processing},
  year={2025},
  volume={33},
  pages={236-250}
}

@article{liu2022neural,
  title={Neural vocoder is all you need for speech super-resolution},
  author={Liu, Haohe and Choi, Woosung and Liu, Xubo and Kong, Qiuqiang and Tian, Qiao and Wang, DeLiang},
  journal={arXiv preprint arXiv:2203.14941},
  year={2022}
}

@inproceedings{chinen2020visqol,
  title={V{iSQOL} v3: An open source production ready objective speech and audio metric},
  author={Chinen, Michael and Lim, Felicia SC and Skoglund, Jan and Gureev, Nikita and O'Gorman, Feargus and Hines, Andrew},
  booktitle={Proc. QoMEX},
  pages={1--6},
  year={2020}
}

@inproceedings{radford2023robust,
  title={Robust speech recognition via large-scale weak supervision},
  author={Radford, Alec and Kim, Jong Wook and Xu, Tao and Brockman, Greg and McLeavey, Christine and Sutskever, Ilya},
  booktitle={Proc. ICML},
  pages={28492--28518},
  year={2023}
}

@inproceedings{abel2018simple,
  title={A simple cepstral domain {DNN} approach to artificial speech bandwidth extension},
  author={Abel, Johannes and Strake, Maximilian and Fingscheidt, Tim},
  booktitle={Proc. ICASSP},
  pages={5469--5473},
  year={2018}
}

@inproceedings{gu2016speech,
  title={Speech bandwidth extension using bottleneck features and deep recurrent neural networks.},
  author={Gu, Yu and Ling, Zhen-Hua and Dai, Li-Rong},
  booktitle={Proc. Interspeech},
  pages={297--301},
  year={2016}
}

@inproceedings{chennoukh2001speech,
  title={Speech enhancement via frequency bandwidth extension using line spectral frequencies},
  author={Chennoukh, Samir and Gerrits, A and Miet, G and Sluijter, R},
  booktitle={Proc. ICASSP},
  pages={665--668},
  year={2001},
}

@INPROCEEDINGS{1326084,
  author={Chen, G. and Parsa, V.},
  booktitle={Proc. ICASSP}, 
  title={{HMM}-based frequency bandwidth extension for speech enhancement using line spectral frequencies}, 
  year={2004},
  volume={1},
  number={},
  pages={I-709}}

@ARTICLE{7078992,
  author={Ling, Zhen-Hua and Kang, Shi-Yin and Zen, Heiga and Senior, Andrew and Schuster, Mike and Qian, Xiao-Jun and Meng, Helen M. and Deng, Li},
  journal={IEEE Signal Processing Magazine}, 
  title={Deep Learning for Acoustic Modeling in Parametric Speech Generation: A systematic review of existing techniques and future trends}, 
  year={2015},
  volume={32},
  number={3},
  pages={35-52},
  doi={10.1109/MSP.2014.2359987}}

@inproceedings{hao2020time,
  title={Time-domain neural network approach for speech bandwidth extension},
  author={Hao, Xiang and Xu, Chenglin and Hou, Nana and Xie, Lei and Chng, Eng Siong and Li, Haizhou},
  booktitle={Proc. ICASSP},
  pages={866--870},
  year={2020}
}

@inproceedings{liu2024audiosr,
  title={AudioSR: Versatile audio super-resolution at scale},
  author={Liu, Haohe and Chen, Ke and Tian, Qiao and Wang, Wenwu and Plumbley, Mark D},
  booktitle={Proc. ICASSP},
  pages={1076--1080},
  year={2024}
}

@inproceedings{shen2018natural,
  title={Natural tts synthesis by conditioning wavenet on mel spectrogram predictions},
  author={Shen, Jonathan and Pang, Ruoming and Weiss, Ron J and Schuster, Mike and Jaitly, Navdeep and Yang, Zongheng and Chen, Zhifeng and Zhang, Yu and Wang, Yuxuan and Skerrv-Ryan, Rj and others},
  booktitle={Proc. ICASSP},
  pages={4779--4783},
  year={2018}
}

@article{yang2025enhancing,
  title={Enhancing Spectrogram Realism in Singing Voice Synthesis via Explicit Bandwidth Extension Prior to Vocoder},
  author={Yang, Runxuan and Li, Kai and Chen, Guo and Hu, Xiaolin},
  journal={arXiv preprint arXiv:2508.01796},
  year={2025}
}

@inproceedings{andreev2023hifi++,
  title={Hifi++: A unified framework for bandwidth extension and speech enhancement},
  author={Andreev, Pavel and Alanov, Aibek and Ivanov, Oleg and Vetrov, Dmitry},
  booktitle={Proc. ICASSP},
  pages={1--5},
  year={2023}
}

@article{liu2022voicefixer,
  title={Voicefixer: A unified framework for high-fidelity speech restoration},
  author={Liu, Haohe and Liu, Xubo and Kong, Qiuqiang and Tian, Qiao and Zhao, Yan and Wang, DeLiang and Huang, Chuanzeng and Wang, Yuxuan},
  journal={arXiv preprint arXiv:2204.05841},
  year={2022}
}

@inproceedings{li2019speech,
  title={Speech Audio Super-Resolution for Speech Recognition.},
  author={Li, Xinyu and Chebiyyam, Venkata and Kirchhoff, Katrin and Amazon, A},
  booktitle={Proc. Interspeech},
  pages={3416--3420},
  year={2019}
}

@inproceedings{buthe2025lightweight,
  title={A lightweight and robust method for blind wideband-to-fullband extension of speech},
  author={B{\"u}the, Jan and Valin, Jean-Marc},
  booktitle={Proc. WASPAA},
  pages={1--5},
  year={2025}
}

@inproceedings{li2021real,
  title={Real-time speech frequency bandwidth extension},
  author={Li, Yunpeng and Tagliasacchi, Marco and Rybakov, Oleg and Ungureanu, Victor and Roblek, Dominik},
  booktitle={Proc. ICASSP},
  pages={691--695},
  year={2021}
}

\end{document}